\documentclass[lettersize, journal]{IEEEtran}
\usepackage[utf8]{inputenc}
\usepackage[T1]{fontenc}
\usepackage{textcomp}
\usepackage{verbatim}
\usepackage{float}
\usepackage{booktabs}
\usepackage{bm}
\usepackage{amsmath}
\usepackage{amssymb}
\usepackage{graphicx}
\usepackage{url}
\usepackage{amsfonts}
\usepackage{times}
\usepackage{graphicx}
\usepackage{latexsym}
\usepackage{dsfont}
\usepackage{amssymb}
\usepackage{amsmath}

\usepackage{verbatim}
\usepackage{subfigure}

\def\bb0{{\mathbb{0}}}

\def\bb{{\mathbf{b}}}

\def\bh{{\mathbf{h}}}

\def\bv{{\mathbf{v}}}
\def\bw{{\mathbf{w}}}

\def\by{{\mathbf{y}}}

\def\b0{{\mathbf{0}}}

\def\bA{{\mathbf{A}}}

\def\bF{{\mathbf{F}}}

\def\bH{{\mathbf{H}}}
\def\bI{{\mathbf{I}}}

\def\bR{{\mathbf{R}}}

\def\sfH{\mathsf{H}}

\def\sf0{{\mathsf{0}}}

\makeatother

\begin{document}
\title{Power consumption and spectral efficiency analysis for uplink analog
radio-over-fiber}
\author{Shehla Amir,~\IEEEmembership{Student,~IEEE,} Miguel Rodrigo Castellanos,~\IEEEmembership{Member,~IEEE,}
Robert W. Heath Jr.,~\IEEEmembership{Fellow,~IEEE} \thanks{S. Amir is with the Department of Electrical and Computer Engineering,
North Carolina State University, Raleigh, NC 27606, USA (email: samir2@ncsu.edu)\protect \\
M.R. Castellanos is with the Department of Electrical Engineering
and Computer Science, University of Tennessee, Knoxville, TN 37996,
USA (email: mrcastellanos@utk.edu)\protect \\
R.W. Heath is with the Department of Electrical and Computer Engineering,
University of California San Diego, La Jolla, CA 92093, USA. (email:
rwheathjr@ucsd.edu). 

This work was supported in part by Samsung Research America and in part by the National Science Foundation under Grant No. NSF ECCS-2414678.}}
\markboth{}{S. Amir \MakeLowercase{\emph{et al.}} }
\maketitle

\begin{abstract}
Radio-over-fiber centralizes radio access networks by using a low-loss optical fiber link between the remote radio head and the central unit. Analog radio-over-fiber (A-RoF) transmits RF signals modulated directly onto an optical carrier, avoiding digitization and digital signal processing at the remote radio head. In this way, A-RoF shifts power-hungry processing from the antenna to the baseband unit. This paper outlines a mathematical framework to analyze the effect of fiber nonlinearity in an uplink wireless system supported by A-RoF. We model an input/output relationship that incorporates the wireless channel, thermal noise, and impairments encountered in the optical fiber channel: chromatic dispersion, electrical-to-optical conversion loss, amplification noise, and fiber nonlinear interference. We compare A-RoF with DSP-assisted A-RoF and digital radio receivers. Our results show that A-RoF achieves higher energy efficiency as compared to digital receivers with 8- and 16-bit analog-to-digital converters and DSP-assisted A-RoF. We further characterize the trade-offs among transmit power, nonlinear interference, and spectral efficiency, demonstrating that nonlinear effects fundamentally limit achievable rates. These results identify the linear operating regions where A-RoF is most effective for uplink wireless communication.
\end{abstract}


\section{Introduction}
 \label{sec:Introduction} 

Centralized radio access networks (C-RANs) shift physical-layer baseband functions from cell sites to a centralized baseband unit (BBU), thereby reducing the complexity of radio units. In digital radio-over-fiber (D-RoF) systems, the remote radio head (RRH) and the BBU are connected using legacy protocols such as the Common Public Radio Interface (CPRI) \cite{CPRISpecificationV7}. CPRI digitizes the radio signal at high resolution for transmission over fiber, but its data-rate demand scales poorly with massive MIMO because the fronthaul load grows rapidly with the number of antennas and the signal bandwidth \cite{Perez-RomeroEtAlTutorialCharacterisationModellingLow2023}. To address these limitations, evolved-CPRI (eCPRI) introduced dynamic functional splits and packet-based fronthaul, offloading part of the baseband processing to the RRH. While some of the eCPRI splits reduce CPRI's data-rate burden, they impose tight latency constraints, can increase RRH power consumption, and complicate synchronization between the RRH and BBU. Furthermore, as RRHs are designed for a specific split, they can become difficult to upgrade \cite{10436913,eCPRI,Perez-RomeroEtAlTutorialCharacterisationModellingLow2023,UdalcovsEtAlTotalCostOwnershipDigital2020}. Moving more digital processing to the RRH also increases the required number of analog-to-digital converters (ADCs). Low-resolution ADCs save power, but they introduce quantization noise that degrades the signal quality \cite{PerryEtAlComparisonAnalogueDigitalFronthaul2020}. Given these limitations for D-RoF systems, analog transmission over fiber is an attractive alternative.

A-RoF simplifies the RRH architecture by relocating baseband processing to the BBU, as originated in C-RAN. Prior work on A-RoF has demonstrated that it can achieve data rates equivalent to, or greater than, CPRI and eCPRI, reaching up to 1 Tb/s \cite{ZhuEtAl102TbCPRIEquivalentRateDirect2022,IshimuraEtAl1032TbCPRIEquivalentRateIFOverFiber2018,ZhuEtAlHighspeedReachextendedIMDD2024,TianEtAl60GHzAnalogRadioOverFiber2017}. These studies have validated the feasibility of high data-rate analog fronthaul, but the analytical insights from the experimental results are limited. An alternate approach is DSP-assisted A-RoF (DA-RoF), where the signal is discretized at the RRH, the number of signal streams reduced, and the result is converted back to analog before fiber transmission \cite{TingEtAlFronthaulOpticalLinksUsing2022,NoorEtAlFlexibleSubcarrierMultiplexingSystem2019}. Recently, hybrid digital/analog RoF (H-RoF) has been introduced, where the signal is converted to discrete time using sigma-delta modulation and an analog residual signal is transmitted alongside for feedback \cite{XuEtAlCoherentDigitalanalogRadiofiberDARoF2022}. The BBU then uses this analog residue to help reconstruct the received RF signal for processing. The choice between these architectures involves a trade-off between baseband complexity, power consumption at the RRH, spectral efficiency, and system flexibility. While standard A-RoF offers higher energy efficiency, H-RoF and DA-RoF can provide greater adaptability and flexibility at RRH for signal transmission \cite{WangEtAlSNRImprovedDigitalAnalogRadioOverFiber2024,WangEtAlSNRImprovedDigitalPCMRadioOverFiber2023}. These differences in A-RoF systems can enable the design of systems tailored to vendor needs, thereby enhancing fronthaul transmission capacity.

Limited work has analytically compared A-RoF with digital fronthaul. For example, \cite{CheAnalogVsDigitalRadioFiber2021} compares A-RoF and D-RoF in terms of spectral efficiency, primarily focusing on quantization noise and its impact on modulation. However, this comparison relies on simplified linearized optical models and does not incorporate nonlinear interference or noise propagation. Rigorous theoretical models developed for long-haul and short-haul fiber transmission systems \cite{ChagnonOpticalCommunicationsShortReach2019,NagarajanEtAlLowPowerDSPBasedTransceivers2021,CarenaEtAlModelingImpactNonlinearPropagation2012,EssiambreTkachCapacityTrendsLimitsOptical2012a,GhozlanKramerModelsInformationRatesMultiuser2017,RamachandranEtAlCapacityRegionBoundsOptical2023} remain underutilized. As a result, a comprehensive analytical framework that incorporates fiber nonlinearities and noise propagation in RoF fronthaul remains an open problem.

In this work, we outline an analytical framework that connects optical-fiber propagation effects with wireless uplink constraints in A-RoF systems. We model both linear and nonlinear fiber impairments, including changes in the refractive index that drive Kerr nonlinearity \cite{AgrawalNonlinearFiberOpticsIts2011}. Such nonlinear effects distort the analog signal, causing self-interference and increasing the effective noise level \cite{AgrellEtAlCapacityNonlinearOpticalChannel2014,SerenaEtAlEnhancedGaussianNoiseModel2020,BeygiEtAlDiscreteTimeModelUncompensatedSingleChannel2012}. In uplink A-RoF, these optical impairments combine with wireless noise introduced at the RRH, further degrading the received signal quality at BBU. The main contributions of this work are summarized below.
 
\textbullet{} We develop a general discrete-time input–output model for the combined optical-wireless uplink channel. Our model combines optical impairments, such as dispersion and nonlinear interference, with wireless effects such as path loss and fading. We model how additive thermal noise introduced at the wireless receiver propagates through the optical link in an A-RoF system, and then extend the model to the DA-RoF system by incorporating quantization noise. This analysis clarifies how signal and noise interact during optical transmission. Furthermore, we use the Gaussian noise model to quantify the impact of nonlinear impairments on A-RoF performance.

\textbullet{} We derive the mutual information between the transmitted signal at the UE and the received signal at the BBU. We leverage it to compare A-RoF with DA-RoF and quantify the performance loss from quantization. Impairments from various optical channel effects, including chromatic dispersion, electrical-to-optical (E/O) converter loss, and nonlinear interference, are combined with impairments from the wireless channel. The results indicate that nonlinearities are not significant for low-power short-reach A-RoF systems. Furthermore, the analysis is extended to receivers with low-resolution ADC systems. For these systems, we assume the fiber link is transparent and only quantization noise is propagated. In this case, A-RoF has higher efficiency as compared to digital systems.

\textbullet{} We compare the power consumption accounting for noise introduced by the RF receiver and power loss due to E/O conversion and amplification. For DA-RoF, the model also includes the ADC power consumption, whereas A-RoF requires less power because no ADC is present at the RRH. The results show that, although A-RoF is affected by nonlinear interference, it achieves higher overall energy efficiency.

The outline of the rest of the paper is as follows. The proposed mathematical system model is discussed in Section \ref{sec:System-model}. In Section \ref{sec:Performance-analysis-of}, the mutual information analysis and power consumption models are outlined. Section \ref{sec:Numerical-analysis} discusses the results from numerical simulations. Finally, Section \ref{sec:Conclusion} concludes the paper.%

\textbf{Notation:} We use the following notation throughout this paper: bold lowercase $\mathbf{a}$ is used to denote column vectors, bold uppercase $\bA$ is used to denote matrices, non-bold letters and $a, A$ are used to denote scalar values. Using this notation, $|a|$ is the magnitude of a scalar, $||\boldsymbol{a}||$ is the $\ell_{2}$ norm, $||\mathbf{a}||_{0}$ is the $\ell_{0}$ norm, $||\boldsymbol{\mathbf{A}}||_{\text{F}}$ is the Frobenius norm, $\text{tr}(\mathbf{A})$ denotes the trace, $\mathbf{A}^{*}$ is the conjugate transpose, $\mathbf{A}^{\text{T}}$ is the matrix transpose, $\mathbf{A}^{-1}$ denotes the inverse of a square matrix, $[\mathbf{a}]_{k}$ is the $k$th entry of $\mathbf{a}$. We use the notation $\mathcal{N_{\mathbb{C}}}(\mathbf{m},\mathbf{R})$ to denote a complex circularly symmetric Gaussian random vector with mean $\mathbf{m}$ and covariance $\mathbf{R}$. We use $\mathbb{E}$ to denote expectation. Convolution is denoted by $*$. The notation $\mathrm{diag}(\mathbf{a})$ denotes a diagonal matrix whose diagonal entries are the elements of $\mathbf{a}$.

\section{System model}

\label{sec:System-model}
In this section, we outline the model for an uplink SIMO system aided by A-RoF. We first present the wireless link and the received signal at the RRH, then model the optical transmission impairments, and finally introduce the A-RoF and DA-RoF signal models. Lastly, we briefly cover the low-resolution digital RF receivers used for comparison in the following sections.

\begin{figure*}[t]
\centering
\includegraphics[width = 0.95\textwidth]{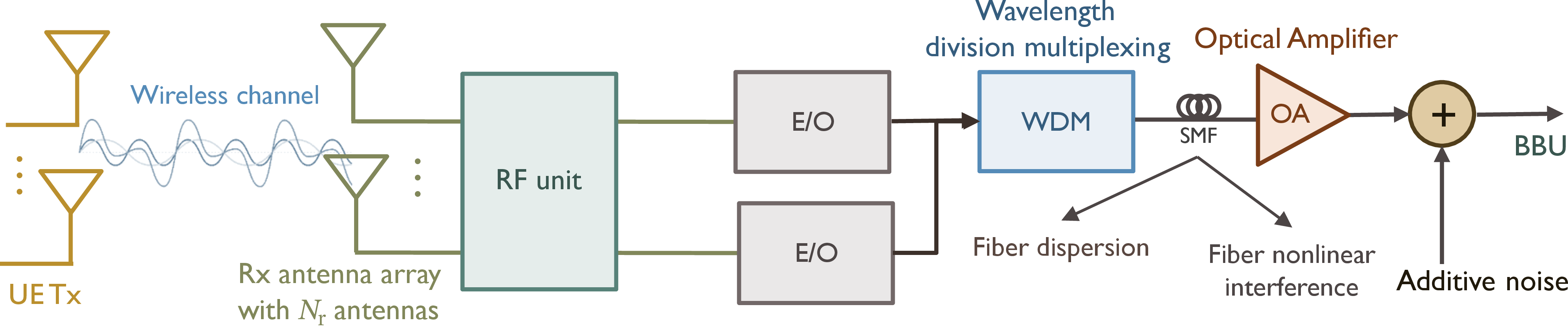}
\caption{Block diagram depicting A-RoF system model. The RRH is equipped with an RF receiver, an RF downconverter, an E/O converter, and a WDM Multiplexer. The transmitted signal goes through the SMF. During optical transmission, the optical signal is amplified and encounters chromatic dispersion. Nonlinear interference and ASE are added to the signal as additive noise. At BBU, the signal is demultiplexed, converted back to an electrical signal, and discretized for baseband processing.
\protect\label{fig:ARoF}}
\end{figure*}

We begin by describing the uplink signal transmitted by the UE. The UE transmits a symbol denoted as $s[n]$. We assume that $s[n]$ is circularly symmetric complex Gaussian with zero mean and unit variance, i.e., $\mathbb{E}[|s[n]|^{2}]=1$. The symbols to be transmitted go through a unit-energy pulse-shaping filter denoted as $g_{\text{tx}}(t)$. The continuous-time transmitted signal is denoted as $x(t)$ with energy $\mathcal{E}{s}$ as $x(t)=\sqrt{\mathcal{E}{s}}\sum_{n=-\infty}^{\infty}s[n] g_{\text{tx}}(t-nT_{\text{w}})$, where the symbol period is $T_{\text{w}} = 1/(2B_{\text{w}}$) and $B_{\text{w}}$ is the signaling bandwidth. At the RRH, the receive antenna array consists of $N_{\mathrm{r}}$ antenna elements, each independently connected to an optical fiber link as shown in Fig.~\ref{fig:ARoF}. We let $\bh_{\text{w}}(t)$ denote the $N_{\text{r}}\times 1$ continuous-time wireless channel vector between the transmitter and the receiver. The thermal noise at the receiver is modeled as an $N_{\text{r}}\times 1$ circularly symmetric complex Gaussian vector $\bv_{\text{w}}(t)$, where $\bv_{\text{w}}(t)\sim\mathcal{N}_{\mathbb{C}}(\mathbf{0},\sigma_{\text{w}}^{2}\bI)$ with variance dependent on Boltzman constant $k_\text{B}$, temperature $T$ and bandwidth as $\sigma_{\text{w}}^{2}=k_{\text{B}}TB_{\text{w}}$. The received continuous-time baseband signal under ideal conditions is
\begin{equation}
\mathbf{y}_{\text{w}}(t)=\bh_{\text{w}}(t)*x(t)+\bv_{\text{w}}(t).\label{eq:pb_rec_sig}
\end{equation}
The received signal in (\ref{eq:pb_rec_sig}) passes through the A-RoF and DA-RoF system to its final destination at the BBU.

\subsection{Optical impairments}

We assume a single-mode WDM fiber link that carries $N_\text{r}$ wavelengths. The $N_\text{r} \times 1$ electrical signal $\by(t)$ is modulated by an E/O converter, mapping its $m$-th component to an optical carrier with wavelength $\lambda_m$, where $m\in \{1,\dots,N\}$. An inline optical amplifier compensates for fiber attenuation. At the BBU, the multiplexed optical signal is converted back to electrical and then digitized for baseband processing.

During fiber transmission, the optical signal is degraded by both linear and nonlinear optical effects. The signal traveling through the fiber experiences linear chromatic dispersion, fiber attenuation, amplification noise, and interference. Nonlinear interference causes interactions of parallel optical carriers as well as changes in signal phase, resulting in distortion that the linear digital receiver cannot compensate \cite{PoggioliniEtAlGNModelFiberNonLinearPropagation2014,SecondiniEtAlNonlinearityMitigationWDMSystems2019,MitraStarkNonlinearLimitsInformationCapacity2001}. This interference arises due to a change in the refractive index of the optical medium as signal power varies \cite{AgrawalFiberopticCommunicationSystems2022,MitraStarkNonlinearLimitsInformationCapacity2001}. The nonlinear interference varies with fiber length, the number of wavelength channels, and core radius. 

Chromatic dispersion occurs due to the waveguide and material dispersion in the fiber \cite{EssiambreEtAlCapacityLimitsOpticalFiber2010}, causing light traveling at different wavelengths to experience distinct group velocities denoted by $\beta_2$ and delays \cite{PlabstEtAlWienerFilterShortReachFiberOptic2020,SavoryDigitalFiltersCoherentOptical2008,liSINRorientedFlexibleQuantization2018}. This effect can be modeled as a frequency selective filter defined in frequency domain for $m$th wavelength and fiber length $L_\text{fiber}$ as $\mathsf{\sfH}_{\text{opt},m}(2\pi f)=e^{-\text{j}\frac{1}{\alpha(\lambda_{m})}(2\pi f)^{2}}$ \cite{SavoryDigitalFiltersCoherentOptical2008,SavoryDigitalCoherentOpticalReceivers2010} where $\alpha(\lambda_{m})=\frac{2\pi}{\pi\beta_{2}(\lambda_{m})L_{\text{fiber}}}$. By taking the inverse Fourier transform of $\mathsf{\boldsymbol{\sfH}}_{\text{opt}}(2\pi f)$, the chromatic dispersion in time domain is 
\begin{align}
\bH_{\text{opt}}(t) & =\text{diag}\Bigg(\sqrt{\alpha(\lambda_{1})}\exp\left(\text{j}\pi\alpha(\lambda_{1})t^{2}\right),\label{eq:chromatic_dispersion-1}\\
 & \dots,\sqrt{\alpha(\lambda_{N_{\text{r}}})}\exp\left(\text{j}\pi\alpha(\lambda_{N_{\text{r}}})t^{2}\right)\Bigg).\nonumber 
\end{align}
Chromatic dispersion introduces different delays and frequency selectivity, thereby adding memory to the system and can be characterized in discrete time by defining $\bH_{\text{opt}}[n]$ as a finite impulse response (FIR) filter with a discrete number of taps \cite{SavoryDigitalFiltersCoherentOptical2008}. In discrete time, the chromatic dispersion represented as an FIR filter has $I$ total taps where $I=|2\pi\beta_{2}(\lambda)B_{\text{w}}^{2}L_{\text{fiber}}|+1$ \cite{ZhongEtAlDigitalSignalProcessingShortReach2018,PlabstEtAlWienerFilterShortReachFiberOptic2020}. For uncompensated short-reach links, where $L_\text{fiber}<100$km, chromatic dispersion is effectively frequency-flat as $I \approx 1$.

The optical amplifier introduces spontaneous emission noise, which enhances the random emission of light. It further amplifies both the signal and noise components of the input by a gain factor denoted as $G_{\text{A}}$. The ASE noise degrades the overall SNR of the optical signal propagating through the fiber \cite{EssiambreEtAlCapacityLimitsOpticalFiber2010}. Each optical carrier experiences different ASE noise, denoted as $v_{\text{ASE},m}(t)$, which is independent of the input signal and modeled as zero mean, circularly symmetric complex additive white Gaussian noise with autocovariance denoted as $\sigma_{\text{ASE},m}^2$. The overall autocovariance of ASE noise is a diagonal matrix denoted as $\bR_{\text{ASE}}$ given by 
$\bR_{\text{ASE}}=\text{diag} (\sigma_{\text{ASE},1}^{2},\ldots,\sigma_{\text{ASE},N}^{2}).$
Each diagonal element of $\bR_{\text{ASE}}$ depends on the amplifier gain $G_{\text{A}}$, spontaneous emission factor $n_{\text{sp}}$, optical frequency $\nu_{m}$, signal bandwidth and Planck's constant $h_{\text{p}}$. The ASE noise variance for $m$th optical channel is \cite{EssiambreEtAlCapacityLimitsOpticalFiber2010}
\begin{equation}
\sigma_{\text{ASE},m}^{2}=B_{\text{w}}n_{\text{sp}}h_{\text{p}}\nu_{m}(G_{\text{A}}-1).
\label{eq:ASEPower-1}
\end{equation}
The ASE noise power increases with amplification gain, which depends on fiber span length and total attenuation \cite{LundbergEtAlPowerConsumptionAnalysisHybrid2017}. Consequently, longer fiber links typically have higher noise power. The noise is also directly proportional to the signal bandwidth. 

The E/O up-converts the baseband signals by modulating them over an optical carrier with signal energy $\mathcal{E}_{\text{opt}}.$ The E/O and O/E converters are lossy components that degrade the signal by a loss factor denoted as $L_{\text{EO}}$ \cite{TianEtAl60GHzAnalogRadioOverFiber2017}. For simplicity, we assume coherent analog optical transmission over the optical fiber \cite{MilovancevEtAlSimplifiedCoherentReceiverAnalogue2021}
and the noiseless received signal at the receiver is $\tilde{\by}(t)=\sqrt{\mathcal{E}_{\text{opt}}L_{\text{EO}}G_{\text{A}}}\bH_{\text{opt}}(t)*\by(t)$. The photodetection process at the O/E introduces additive shot noise, denoted as $\bv_{\text{shot}}(t)$, which is typically insignificant as compared to ASE noise \cite{EssiambreEtAlCapacityLimitsOpticalFiber2010}. Thus, ASE noise and nonlinear interference are the dominant noise sources in the optical link.

Nonlinear interference (NLI) is another major impairment in the analog optical link, alongside chromatic dispersion and ASE noise. It arises from intensity-dependent changes in the fiber's refractive index, which cause waveform distortions and random mixing among WDM channels which very with total signal power over the fiber link \cite{EssiambreEtAlCapacityLimitsOpticalFiber2010,PoggioliniEtAlGNModelFiberNonLinearPropagation2014}. Although NLI is often considered negligible in short-reach links, it becomes nontrivial in A-RoF because the uplink signal already contains wireless noise, and both distortions interact during propagation. Following the standard Gaussian Noise (GN) modeling framework, we assume that nonlinear distortions become sufficiently randomized to be statistically independent of the transmitted waveform \cite{PoggioliniEtAlGNModelFiberNonLinearPropagation2014}. Under this widely used assumption, the accumulated nonlinear distortion can be treated as an additive Gaussian noise, which is a conservative approximation of the nonlinear fiber channel.


The nonlinear interference coefficient for single polarization depends on the signal shape and system parameters derived from the NLSE, such as the nonlinear coefficient $\gamma$, attenuation, the system bandwidth, and the group velocity dispersion coefficient. The distortion vector denoted as $\mathbf{v}_{\text{NLI}}(t)$ has autocovariance $\bR_{\text{NLI}}(\tau)$ and follows a complex Gaussian distribution $\mathcal{N}_{\mathbb{C}}\sim(\mathbf{0},\mathbf{R}_{\text{NLI}}(\tau))$. The autocovariance of $\bv_{\text{NLI}}(t)$, denoted by $\bR_{\text{NLI}}(\tau)$, depends on the nonlinear interference coefficient $\eta(\tau)$ and the autocovariance of the input signal $\bR_{\tilde{\by}}(\tau)=\mathbb{E}[\tilde{\by}(t)\tilde{\by}(t+\tau)]$.This variance is explicitly derived following the perturbative approach in \cite{SecondiniEtAlNonlinearityMitigationWDMSystems2019,EllisEtAlApproachingNonLinearShannonLimit2010,PoggioliniEtAlGNModelFiberNonLinearPropagation2014}. Assuming a wide-sense stationary signal propagating over fiber, the autocovariance matrix is
\begin{equation}
\bR_{\text{NLI}}(\tau)=\eta(\tau)\bR_{\tilde{\by}}(\tau)\bR_{\tilde{\by}}(\tau)\bR_{\tilde{\by}}(\tau).\label{eq:nonlinearilty-1}
\end{equation}
The nonlinear interference coefficient in (\ref{eq:nonlinearilty-1}) increases with the received signal power. It can be further increased by using a larger number of WDM channels and varying channel spacings. 

Combining the above impairments with the frequency-flat optical dispersion and loss over fiber. The received signal denoted as $\bw(t)$ after demultiplexing is
\begin{align}
\bw(t) & =\sqrt{\mathcal{E}_{\text{opt}}L_{\text{EO}}G_{\text{A}}}\bH_{\text{opt}}(t)\tilde{\by}(t)
 +\bv_{\text{NLI}}(t)+\bv_{\text{ASE}}(t).
 \label{eq:OpticalRecSignal-1} 
\end{align}
 Nonlinear interference becomes significant as the number of independent channels increases and the wavelength spacing decreases. The nonlinear interference is modeled as additive Gaussian noise with power given in (\ref{eq:nonlinearilty-1}).
\begin{figure*}[t]
\centering
\includegraphics[width = 0.95\textwidth]{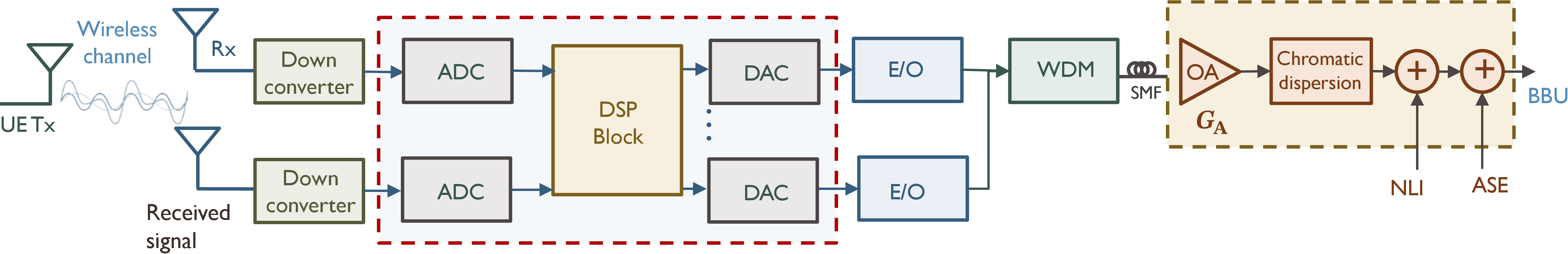}
\caption{Block diagram depicting DA-RoF system model. The RRH is equipped with an RF receiver and downconverter, an ADC, a DAC, a resampling block, an E/O converter, and a WDM Multiplexer. The received signal is converted to digital and resampled for optical transmission. The transmitted signal goes through the fiber optic channel, where it is amplified and encounters chromatic dispersion. Nonlinear interference and ASE are added to the signal as additive noise. At BBU, the signal is demultiplexed and converted back to electrical for baseband processing. \protect\label{fig:DARoF}}
\end{figure*}

\subsection{Analog radio-over-fiber}

The RRH architecture, depicted in Fig. \ref{fig:ARoF}, comprises receive antennas, downconverters, E/O converters, a wavelength division multiplexer, and an SMF connecting the RRH to the BBU. The received electrical signal at each antenna $\by_{\text{w}}(t)$ is passed through the E/O converters, which modulates each received signal onto an optical carrier at wavelength $\lambda_{m}$ where $m\in\{1,...,N_{\text{r}}\}$. The signals are then multiplexed on an SMF using a WDM, which enables transmission of $N_{\text{r}}$ distinct signals to the BBU. At the BBU, the analog signal undergoes O/E conversion and is subsequently digitized using a low-loss ADC, effectively adding negligible quantization noise. We assume optimal signal combining at the BBU and focus primarily on impairments of the analog optical-wireless channel. 

In the A-RoF system, the analog received signal passes through the optical system as outlined above. For our $10$ km link, chromatic dispersion is small where $I\approx1$ and can be digitally compensated at the BBU \cite{SavoryDigitalFiltersCoherentOptical2008}. Therefore we can treat $\bH_\text{opt}\approx \bI$. The baseband signal is discretized by passing it through an ADC and a receiver matched filter. The signal is sampled at a time interval $T$, and the discrete A-RoF signal is given by $\bw_{\text{A-RoF}}[n]=\bw_{\text{A-RoF}}(nT)$. The frequency-selective wireless channel is modeled as an FIR  filter with $A$ filter taps. The discrete time baseband signal at the BBU is 
\begin{eqnarray}
\mathbf{w}_{\text{A-RoF}}[n] & = & \sqrt{L_{\text{EO}}G_{\text{A}}\mathcal{E}_{\text{opt}}}\sum_{a=0}^{A-1}\bh_{\text{w}}[a]x[n-a]\nonumber \\
 &  & +\sqrt{L_{\text{EO}}G_{\text{A}}\mathcal{E}_{\text{opt}}}\bv_{\text{w}}[n]\label{eq:A_RoF_Final_eq}\\
 &  & +\bv_{\text{NLI}}[n]+\bv_{\text{ASE}}[n].\nonumber 
\end{eqnarray}
The signal at the BBU has a bandwidth $B_{\text{w}}$, and the additive noise in the system is also bandlimited. The overall noise of system denoted as $\bv_{\text{A-RoF}}[n]$ the system is
\begin{equation}
\bv_{\text{A-RoF}}[n]=\sqrt{L_{\text{EO}}G_{\text{A}}\mathcal{E}_{\text{opt}}}\bv_{\text{w}}[n]+\bv_{\text{NLI}}[n]+\bv_{\text{ASE}}[n].
\end{equation}
The total noise in the signal depends on the system impairments. The additive noise covariance depends on the optical and power of the received wireless signal at the RRH, as 
\begin{align}
\bR_{\bv_{\text{A-RoF}}}[k] & =\mathbb{E}[\bv_{\text{A-RoF}}[n]\bv_{\text{A-RoF}}^{*}[n+k]]\nonumber \\
 & =L_{\text{EO}}G_{\text{A}}\mathcal{E}_{\text{opt}}\sigma_{\text{w}}^{2}\bI+\bR_{\text{NLI}}[k]+\bR_{\text{ASE}}.
\end{align}
Thus, changing the optical power or controlling fiber impairments will affect the received signal power.%

\subsection{Digital-Analog radio-over-fiber}

DA-RoF systems process received signals at the RRH, converting them back to analog form for optical transmission. At the RRH, the received wireless baseband signal is quantized using low-resolution ADCs, thus, the received wireless signal experiences quantization distortion. We model the ADC output using the additive quantization noise model (AQNM), which expresses the quantized signal as a scaled version of the input plus uncorrelated quantization noise \cite{ZhangEtAlCellFreeMassiveMIMOFewbit2020a,OrhanEtAlLowPowerAnalogtodigitalConversion2015}. With this model, the quantized signal $\mathbf{y}_{\mathrm{q}}[n]$ is digitally precoded by the matrix $\mathbf{F}_{\mathrm{r}}$, resampled, and converted back into analog optical signals. These signals are finally mapped onto $N_{\mathrm{s}}$ WDM channels for fiber transmission. The modulated signals are then multiplexed using WDM for transmission over fiber. In this case, $N_{\text{s}}$ optical channels are required for optical transmission.

At the RRH, the ADC distorts the signal by a quantization coefficient $\epsilon_{\text{ADC}}$ and introduces additive quantization noise denoted as $\bv_{\text{ADC}}[n]$. Following the assumption of AQNM, the quantization noise by the ADC is drawn from a circularly symmetric complex Gaussian distribution with zero mean, and covariance denoted as $\bR_{\text{ADC}}[n]$. The discrete time output of the ADC then depends on the quantization coefficient $\epsilon_{\text{ADC}}$ and additive noise $\bv_{\text{ADC}}[n]$ as 
\begin{equation}
\by_{\text{q}}[n]=\epsilon_{\text{ADC}}\by_{\text{w}}[n]+\bv_{\text{ADC}}[n].\label{eq:Quantization-1}
\end{equation}
The quantization coefficient $\epsilon_{\text{ADC}}$depends on quantization bits $b_{\text{ADC}}$ as \cite{OrhanEtAlLowPowerAnalogtodigitalConversion2015} 
\begin{equation}
\epsilon_{\text{ADC}}=1-\frac{\pi\sqrt{3}}{2b_{\text{ADC}}^{2}}.
\end{equation}
The quantization noise is assumed to be uncorrelated to the received signal $\by_{\text{w}}[n]$. The $N_{\text{r}}\times N_{\text{r}}$ covariance matrix of the received signal at RRH is denoted by $\mathbf{R}_{\mathbf{\by}}[k]$, where each diagonal entry contains the variance of the received baseband signal for each RF chain. The covariance matrix of the quantization noise denoted as $\mathbf{R_{\text{ADC}}}[n]$ can then be expressed as
\begin{equation}
\bR_{\text{ADC}}[k] =\epsilon_{\text{ADC}}(1-\epsilon_{\text{ADC}})\mathbb{E}[\by_{\text{w}}[n]\by_{\text{w}}^{*}[n+k]].
\end{equation}
The additive noise introduced by the quantizer propagates through the optical channel. %

The DA-RoF system can be considered a radio relay system to improve the performance of the over the A-RoF system. The RRH acts as a MIMO relay, combining and beamforming the signal for transmission over an optical link. One example of this application is that, if channel information is available, MRC for combining and MRT for transmission over an optical link can be used. The combined system is then denoted by $N_{s}\times N_{\text{r}}$ $\bF_{\text{r}}$. The output of the RRH combiner is 
\begin{equation}
\mathbf{y}_{\text{r}}[n]=\bF_{\text{r}}\by_{\text{q}}[n].
\end{equation}
The RRH signal is then converted to a continuous time analog signal for transmission over the optical link. We assume the DAC adds jitter noise denoted as $\bv_{\text{DAC}}(t)$ in the signal. Using AQNM, the jitter noise is also additive and follows a complex Gaussian distribution \cite{XiongEtAlPerformanceAnalysisMassiveMIMO2021a}. As the signal passes through the DAC, another quantization error is introduced using a similar technique to AQNM. Both ADC quantization noise and DAC jitter noise are drawn from a complex Gaussian random distribution \cite{OrhanEtAlLowPowerAnalogtodigitalConversion2015}. The DAC noise just adds further distortion in the case of lower resolution DACs. The DAC coefficient is defined similarly as ADC where $\epsilon_{\text{DAC}}=1-\frac{\pi\sqrt{3}}{2b_{\text{DAC}}^{2}}$ \cite{XiongEtAlPerformanceAnalysisMassiveMIMO2021a}. 

Similar to A-RoF, the signal traveling in the DA-RoF link experiences linear chromatic dispersion and nonlinear interference as outlined above. In the case of DA-RoF, the resampling bandwidth affects the variance of the AWGN noise in the system and is therefore another parameter that needs to be accounted for in the system. The optical signal then travels along the optical channel, encountering similar impairments, such as ASE and shot noise, as discussed for A-RoF. The received signal at BBU, denoted as $\bw_{\text{DA-RoF}}(t)$, goes through an optical detector and then through ADC, and the final received signal at the BBU. 

For DA-RoF, the received signal at the BBU is
\begin{align}
\bw_{\text{DA-RoF}}[n]&=\sqrt{L_{\text{EO}}G_{\text{A}}\mathcal{E}_{\text{opt}}\epsilon_{\text{ADC}}\epsilon_{\text{DAC}}} \bF_{\text{r}}\bh_{\text{w}}x[n] \\ &+\bv_{\text{DA-RoF}}[n] \nonumber.
\end{align}
The total noise in the DA-RoF system is
\begin{flalign}
\bv_{\text{DA-RoF}}[n] & =\sqrt{L_{\text{EO}}G_{\text{A}}\mathcal{E}_{\text{opt}}\epsilon_{\text{ADC}}\epsilon_{\text{DAC}}} \bF_{\text{r}}\bv_{\text{w}}[n]\\
 & +\sqrt{L_{\text{EO}}G_{\text{A}}G_{\text{A}}\epsilon_{\text{DAC}}} \bF_{\text{r}}\bv_{\text{ADC}}[n]\nonumber \\
 & +\sqrt{L_{\text{EO}}G_{\text{A}}G_{\text{A}}\epsilon_{\text{DAC}}} \bv_{\text{DAC}}[n]\nonumber \\
 & +\bv_{\text{NLI}}[n]+\bv_{\text{ASE}}[n].\nonumber 
\end{flalign}
The $N_{\text{s}}\times N_{\text{s}}$ autocovariance matrix for the noise denoted as $\bR_{\bv_{\text{DA-RoF}}}[k]$ is used to estimate the mutual information.

\subsection{Digital RF receiver}

In this system, the received signal is quantized at RRH, and the digital signal is transmitted over fiber. In this case, we assume that the signal transmitted over fiber is optimal \cite{Perez-RomeroEtAlTutorialCharacterisationModellingLow2023}. To focus the shift on power consumption at RRH, we assume that the ADCs are lossy and add quantization noise. The incoming baseband wireless signal is quantized and denoted by $\by_{\text{q}}[n]$. At the RRH, the ADC distorts the signal by a quantization coefficient $\epsilon_{\text{ADC}}$ and introduces additive quantization noise denoted as $\bv_{\text{ADC}}[n]$. Following the assumption of AQNM, the quantization noise by the ADC is drawn from a circularly symmetric complex Gaussian distribution with zero mean and covariance denoted as $\bR_{\text{ADC}}[n]$. The discrete time output of the ADC then depends on the quantization coefficient $\epsilon_{\text{ADC}}$
and additive noise $\bv_{\text{ADC}}[n]$ as 
\begin{align}
\by_{\text{q}}[n] & =\epsilon_{\text{ADC}}\by_{\text{w}}[n]+\bv_{\text{ADC}}[n]\label{eq:Quantization}\\
 & =\epsilon_{\text{ADC}}\left(\sum_{a=0}^{A-1}\bh_{\text{w}}[a]x[n-a]+\bv_{\text{w}}[n]\right)+\bv_{\text{ADC}}[n].
\end{align}
The quantization noise is assumed to be uncorrelated to the received signal $\by_{\text{w}}[n]$. The $N_{\text{r}}\times N_{\text{r}}$ covariance matrix of the received signal at RRH is denoted by $\mathbf{R}_{\mathbf{\by}}[k]$, where each diagonal entry contains the variance of the received baseband signal for each RF chain. The covariance matrix of the quantization noise denoted as $\mathbf{R_{\text{ADC}}}[n]$ can then be expressed as $\bR_{\text{ADC}}[k] = \epsilon_{\text{ADC}}(1-\epsilon_{\text{ADC}})\mathbb{E}[\by_{\text{w}}[n]\by_{\text{w}}^{*}[n+k]]$. The additive noise introduced by the quantizer propagates through the optical channel.

\section{Mutual information and power consumption analysis} \label{sec:Performance-analysis-of}

This Section analyzes the spectral efficiency and power consumption of A-RoF and DA-RoF systems. We first derive the mutual information for both architectures, accounting for impairments in the optical and wireless channels. We further model the power consumption of both systems, taking into account the different system blocks and their respective impacts.

\subsection{Mutual information}
For simplification and to focus on the effect of noise impairments, we assume the wireless channel is also frequency-flat. For A-RoF, the received signal at BBU contains wireless additive thermal noise, ASE, and nonlinear interference. The mutual information is
\begin{equation}
\mathcal{I}_{\text{A-RoF}}=\text{log}_{2}\left(\det\left(\bI+\mathcal{E}_{s}\mathcal{E}_{\text{opt}}G_{\text{A}}L_{\text{EO}}\boldsymbol{\bh}_{\text{w}}\bR_{\text{A-RoF}}^{-1}\boldsymbol{\bh}_{\text{w}}^{*}\right)\right).
\label{eq:MIAROF}
\end{equation}
It is not explicitly clear from (\ref{eq:MIAROF}) the inverse relationship between signal power and mutual information. However, for uncorrelated channels and the received signal, we obtain an explicit relationship that allows us to directly observe the inverse relationship \cite{AgrellEtAlInformationtheoryfriendlyModelsFiberopticChannels2015}. 

In contrast, the mutual information for the DA-RoF system includes the additional impact of quantization noise due to ADC and DAC. The mutual information in this case is 
\begin{align}
\mathcal{I}_{\text{DA-RoF}} & =\text{log}_{2}\bigg(\det\bigg(\bI+\\
&\mathcal{E}_{s}\mathcal{E}_{\text{opt}}G_{\text{A}}L_{\text{EO}}\epsilon_{\text{ADC}}\epsilon_{\text{DAC}}\bF_{\text{r}}\bh_{\text{w}}\bR_{\text{DA-RoF}}^{-1}\bh_{\text{w}}^{*}\bF_{\text{r}}^{*}\bigg)\bigg).\nonumber 
\end{align}
Since $\epsilon_{\text{ADC}}$ and $\epsilon_{\text{DAC}}$ are both less than or equal to one, their squared product can act as a multiplicative penalty on the received power. As a result, even under ideal channel conditions, coarse quantization at the RRH results in a reduction of the overall achievable rate. In DA-RoF, this effect is compounded because the quantization noise is also transported over the fiber, thus, the mutual information depends not only on the fiber and wireless impairments but also on impairments introduced by ADC/DAC resolution. At low ADC/DAC resolution, the quantization noise power increases and is forwarded to the BBU, so the DA-RoF link effectively carries more noise than the A-RoF link, further limiting the information rate.

For the baseline digital RF receiver system, we consider a simple input/output system impacted by quantization noise. The total noise term in this system, denoted by $\mathbf{v}_{\text{D-RF}}[n]$, includes both the thermal noise from the RF front-end as well as the quantization noise introduced by the ADCs. The total noise is 
\begin{equation}
\bv_{\text{D-RF}}[n]=\sqrt{\epsilon_{\text{ADC}}}\bv_{\text{w}}[n]+\bv_{\text{ADC}}[n].
\end{equation}

The noise autocovariance is $\mathbf{R}_{\text{D-RF}}[k]=\mathbb{E}[\mathbf{v}_{\text{D-RF}}[n]\mathbf{v^{*}_{\text{D-RF}}}[n+k]].$ The mutual information for this system is 
\begin{equation}
\mathcal{I}_{\text{D-RF}}=\text{log}_{2}\bigg(\det\bigg(\bI+\mathcal{E}_{s}\epsilon_{\text{ADC}}\bh_{\text{w}}\bR_{\text{D-RF}}^{-1}\bh_{\text{w}}^{*}\bigg)\bigg).
\end{equation}
The mutual information is used to compare with RoF systems under identical wireless channel conditions and system parameters.

\subsection{Power consumption} \label{sec:Power-consumption-and}

We develop RRH-side power models for A-RoF, DA-RoF, and digital RF systems to compare spectral efficiency fairly and account for power losses. The A-RoF-aided RRH comprises power-consuming elements, such as an E/O converter, low-noise amplifiers (LNAs), and an optical amplifier. The optical amplifier is used to amplify the signal for RoF systems, thereby increasing the received signal at the antenna. This signal is amplified using an LNA, which consumes power denoted by $P_{\text{RF}}$. The E/O converter modulates the RF signal to an optical signal generated using a laser source. The input laser has power $P_{\text{in}}$. The signal experiences conversion loss at the E/O interface, denoted as $L_{\text{EO,dB}}$. The loss is due to the beating of the oscillator, which results in some power being lost in the system \cite{EssiambreEtAlCapacityLimitsOpticalFiber2010}. 

The fiber, due to its intrinsic properties, attenuates the propagating signal. The attenuation is quantified using the fiber attenuation coefficient, denoted by $\alpha$, and the transmission distance, $z$. An amplifier with gain denoted $G_{\text{A}}$ is used to counter the attenuation effect of the fiber. The amplifier gain is set to compensate for the fiber attenuation, calculated as $G_{\text{A}}=e^{\alpha z}$. In this work, we consider Erbium-doped fiber amplifiers (EDFAs) used to amplify the signal over SMF. The power consumption of EDFA, denoted by $P_{\text{A}}$, depends on the amplifier power conversion ratio denoted as $\eta_{\text{pc}}$ and number of WDM channels \cite{LundbergEtAlPowerConsumptionAnalysisHybrid2017}. The signal for each WDM channel has an input power of $P_{\text{in}}$. For $N_{\text{r}}$ channels with amplifier gain $G_{\text{A}}$ the total power consumed by amplifier is \cite{LundbergEtAlPowerConsumptionAnalysisHybrid2017} 
\begin{equation}
P_{\text{A}}=\frac{1}{\eta_{\text{pc}}}N_{\text{r}}P_{\text{laser}}\left(1-\frac{1}{G_{\text{A}}}\right).\label{eqn:EDFA}
\end{equation}
The amplification gain scales with the input power of the channel and the number of antenna elements. The total power consumed at the A-RoF-aded RRH then depends on the amplifier's power, input optical power, power consumed by E/O to counter loss, $P_{\text{RF}}$, and the number of antenna elements. Then the total power denoted as $P_{\text{A-RoF}}$ is 
\begin{equation}
P_{\text{A-RoF}}=N_{\text{r}}(P_{\text{laser}}+P_{\text{RF}})+P_{\text{A}}+P_{\text{WDM}}.
\end{equation}
The overall power consumption for A-RoF enabled using WDM scales linearly with the number of channels. Thus, increasing the number of antennas also increases power consumption. 

\begin{table}[t]
\centering
\setlength{\tabcolsep}{6pt} 
\renewcommand{\arraystretch}{1} 
\caption{\bf System parameters for simulations}
\begin{tabular}{lll}
\hline
Symbol & Value & Description \\
\hline
$\alpha$ & 0.2 dB/km & Fiber attenuation\\
$\beta_{2}$ & -21.7 $\text{ps}^{2}/\text{km}$ & Fiber dispersion at $1550$ nm\\
$h$ & $6.62\times 10^{-34}$ $\text{J}/\text{Hz}$ & planck's constant\\
$\gamma$ & 1.27 $\text{W}^{-1}\text{km}^{-1}$ & Nonlinear coefficient\\
$G_{\text{A}}$ & 25 dB & Amplification gain\\
$k_{\text{e}}$ & 1.23$\times$$10^{-23}$ J/K & Boltzmann\textquoteright s constant\\
$T$ & 290 K & Temperature\\
$B_{\text{w}}$ & 100 MHz & Signal bandwidth\\
$B_{\text{opt}}$ & 50 GHz & WDM spacing\\
$N_{\text{r}}$ & 4,16,64, 128 & Number of antennas\\
$P_{\text{RF}}$ & 56 mW & RF power\\
$P_{\text{WDM}}$ & 22 W & WDM power\\
$P_{\text{laser}}$ & 10 dBm & Laser power\\
$\eta_{\text{sp}}$ & 1.58 & Spontaneous emission factor\\
$\lambda$ & 1550 nm & WDM wavelength\\
$L_{\text{fiber}}$ & 10 km & Fiber length\\
$c$ & $3\times10^{8}$$\text{m}/\text{s}$ & Speed of light\\
$L_{\text{EO}}$ & -7 dB & E/O conversion loss\\
\end{tabular}
\label{table:SysParam}%
\end{table}

DA-RoF transmits a digital signal over the fiber channel, therefore, an ADC is required at the RRH. The received electrical signal is passed through the ADC to convert into a digital signal. The power is then dependent on the bandwidth of the signal $B$, number of quantization bins $b$, and ADC conversion energy $\eta_{\text{ADC}}$ \cite{OrhanEtAlLowPowerAnalogtodigitalConversion2015}. The power consumed by the ADC denoted as $P_{\text{ADC}}$ is
\begin{equation}
P_{\text{ADC}}=2^{b_{\text{ADC}}}B_{\text{w}}\eta_{\text{ADC}}.
\end{equation}
The total power consumed at the RRH, denoted as$P_{\text{DA-RoF}}$, is dependent on the power consumed by the ADC, E/O converters, electrical antenna elements, and amplifier. The E/O converters add their conversion loss and insertion loss. An EDFA amplifier is assumed for DA-RoF as covered in (\ref{eqn:EDFA}). Let $P_{\text{DAC}}$ be the power consumed by the digital-to-analog converter, $P_{\text{laser}}$ the power required by the optical laser source, and $P_{\text{DSP}}$ the power overhead from clocking, synchronization, and baseband logic. Then, for $N_{\text{r}}$ receive chains, the total power is modeled as the total power consumed by the RRH for DA-RoF, which is 
\begin{equation}
P_{\text{DA-RoF}}=N_{\text{r}}(P_{\text{ADC}}+P_{\text{DAC}}+P_{\text{laser}}+P_{\text{RF}})+P_{\text{WDM}}+P_{\text{A}}+P_{\text{DSP}}.
\end{equation}
DA-RoF inherently consumes more power than its analog counterpart, but this can be alleviated by using low-resolution ADCs.

We then compare the power consumption at RRH for the digital RRH system. We assume that the main contributors to power consumption for an electrical receiver are LNAs, mixers, and ADCs. The CPRI interface, responsible for transporting the digitized signal to the baseband unit, consumes power denoted as $P_{\text{CPRI}}$. The power of the RF chain is scaled by the number of receive antennas, and then the total power consumed by the RRH is
\begin{equation}
P_{\text{D-RF}}=N_{\text{r}}(2P_{\text{ADC}}+P_{\text{RF}})+P_{\text{CPRI}}.\label{eq:PRF}
\end{equation}
As $P_{\text{RF}}$ is fixed, the power consumption is dependent on the ADCs in (\ref{eq:PRF}), which drive the power consumption in this case.

\begin{figure}[t]
    \centering
    \includegraphics[width = \columnwidth]{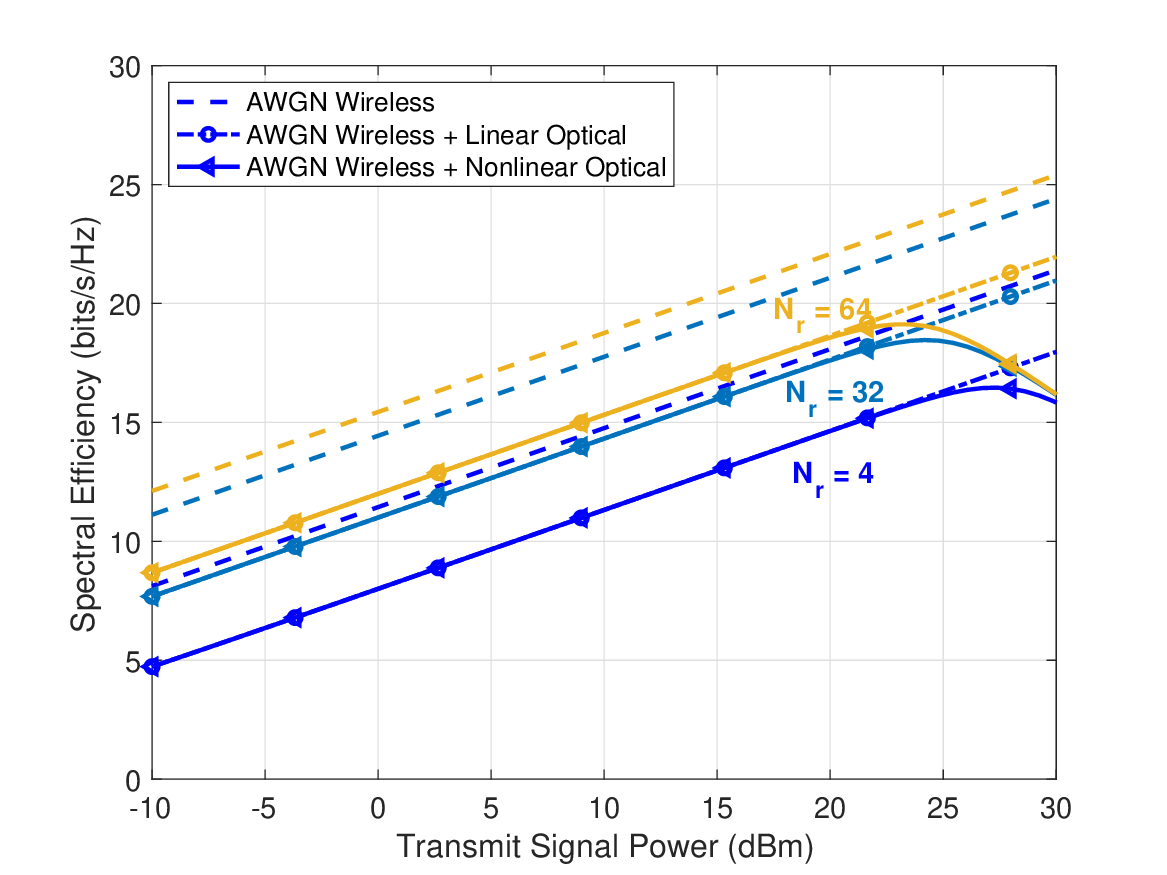}
    \caption{Comparing linear and nonlinear antennas with an ideal AWGN system. The RoF system is limited by nonlinear interference as the transmitted power increases.}
    \label{fig:differentRoFmodels-1}
\end{figure}

\begin{figure}[t]
\centering

 \includegraphics[width = \columnwidth]{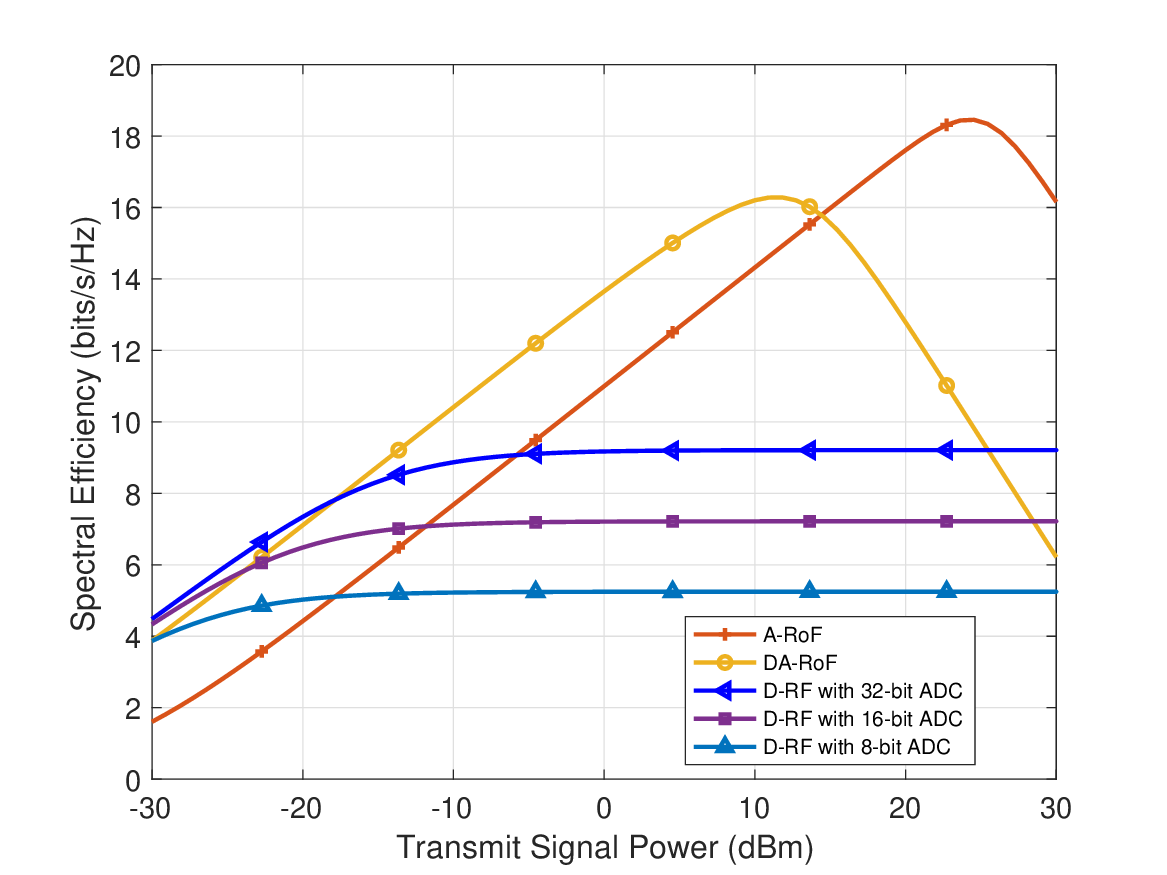}
\caption{Comparing RoF systems with no RoF system. At higher powers, A-RoF systems outperform their digital counterparts. However, at lower transmission powers, DA-RoF achieves higher spectral efficiency.\protect\label{fig:differentRoFmodels-2}}
\end{figure}

\begin{figure}[t]
\centering
\includegraphics[width = \columnwidth]{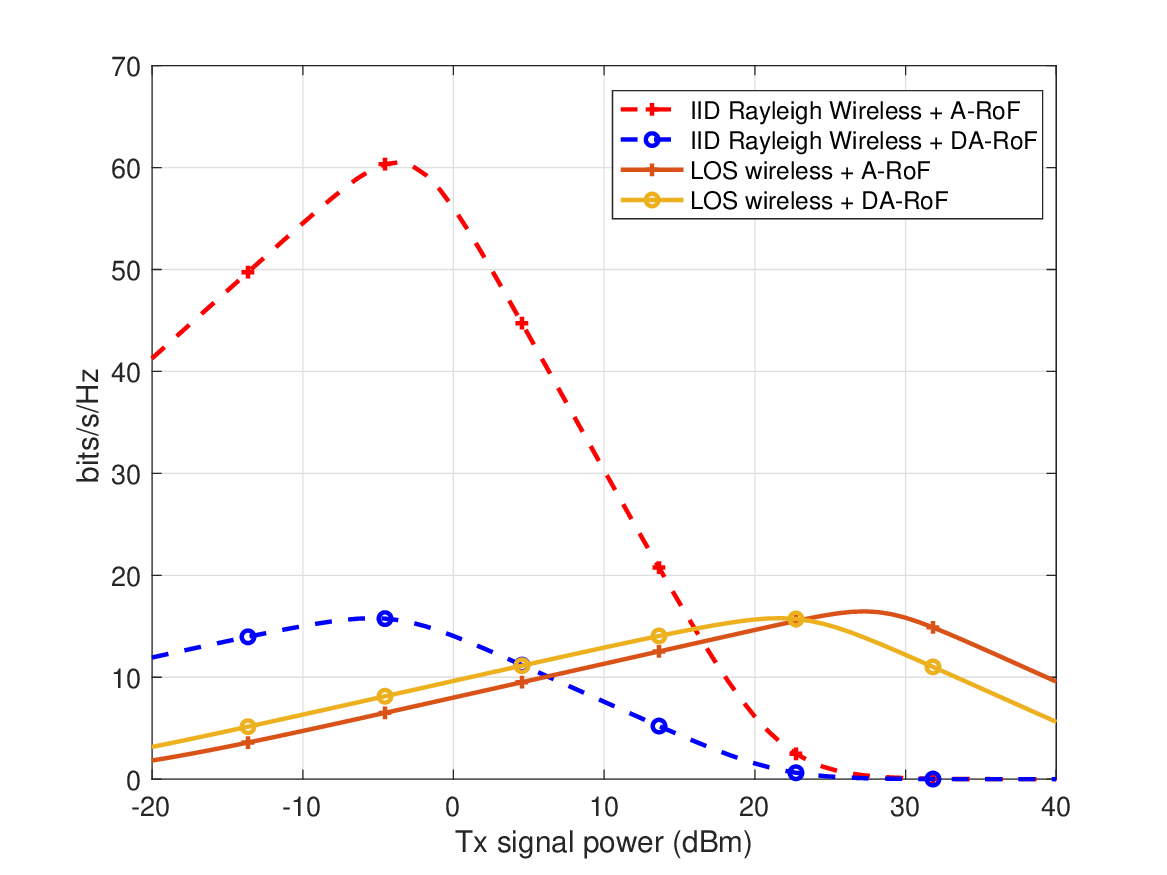}

\caption{Comparing the effect of fading in RoF systems. Wireless fading affects the optical transmission, leading to nonlinear interference at lower transmission powers.\protect\label{fig:DifferentFading}}
\end{figure}

\section{Numerical analysis} \label{sec:Numerical-analysis}

\begin{figure}[!t]
\raggedright
\includegraphics[width = \columnwidth]{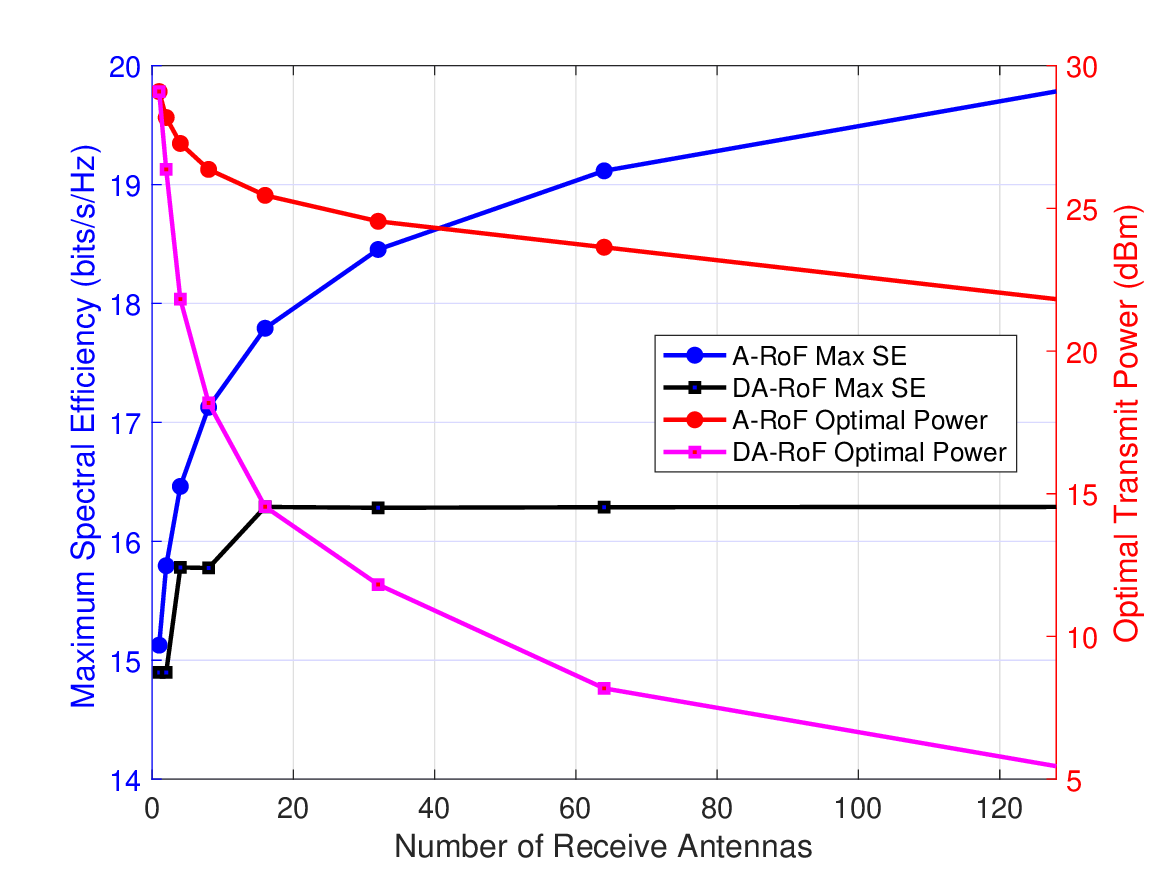}\caption{Comparing the maximum spectral efficiency and its respective peak transmission power against the number of antennas. As the number of antennas increases, the power at which nonlinear interference occurs reduces.
\protect\label{fig:differentRoFmodels}}
\end{figure}

\begin{figure}[t]
\centering
\includegraphics[width = \columnwidth]{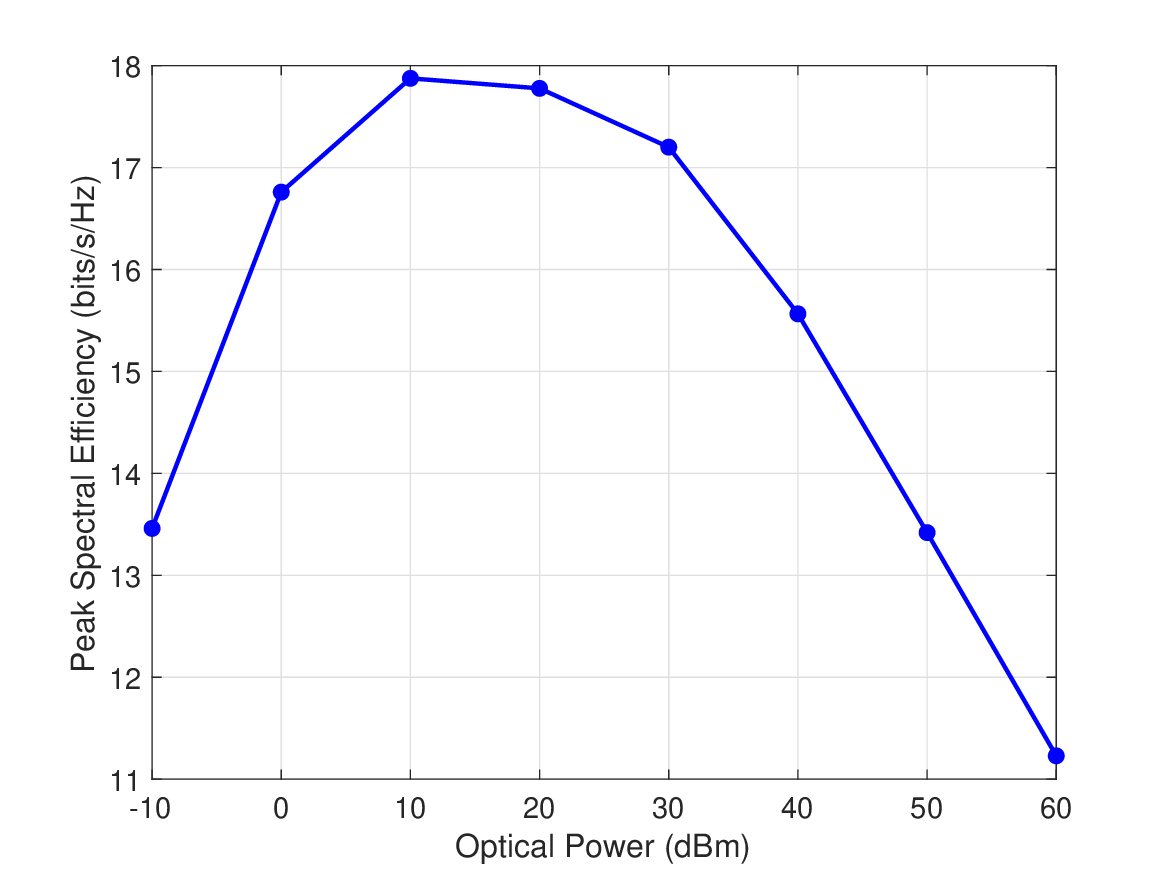}

\caption{Comparing the peak spectral efficiency for the A-RoF system against varying optical powers. At around 10-20 dBm, the rate is high, and then it slowly reduces with an increase in optical power over the fiber. \protect\label{fig:Comparingtheopt}}
\end{figure}

\begin{figure}[t]
\centering
\includegraphics[width=\columnwidth]{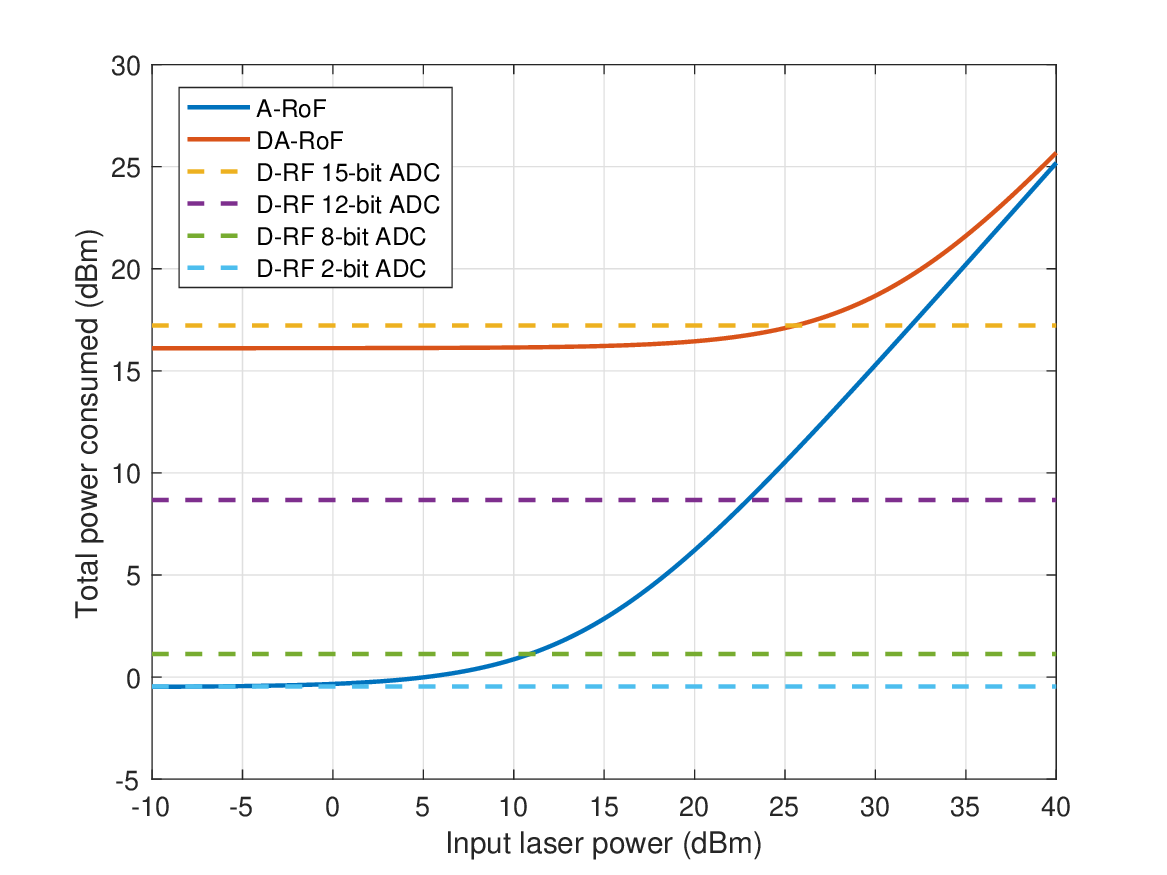}

\caption{Comparing the power consumed by the RoF systems. The DA-RoF system consumes the highest power, followed by the D-RF system, as the number of bits increases. A-RoF consumes power compared to the D-RF system with low-resolution ADCs.} \protect\label{fig:Comparingthepower}

\end{figure}

\begin{figure}[t]
\raggedright
\includegraphics[width = \columnwidth]{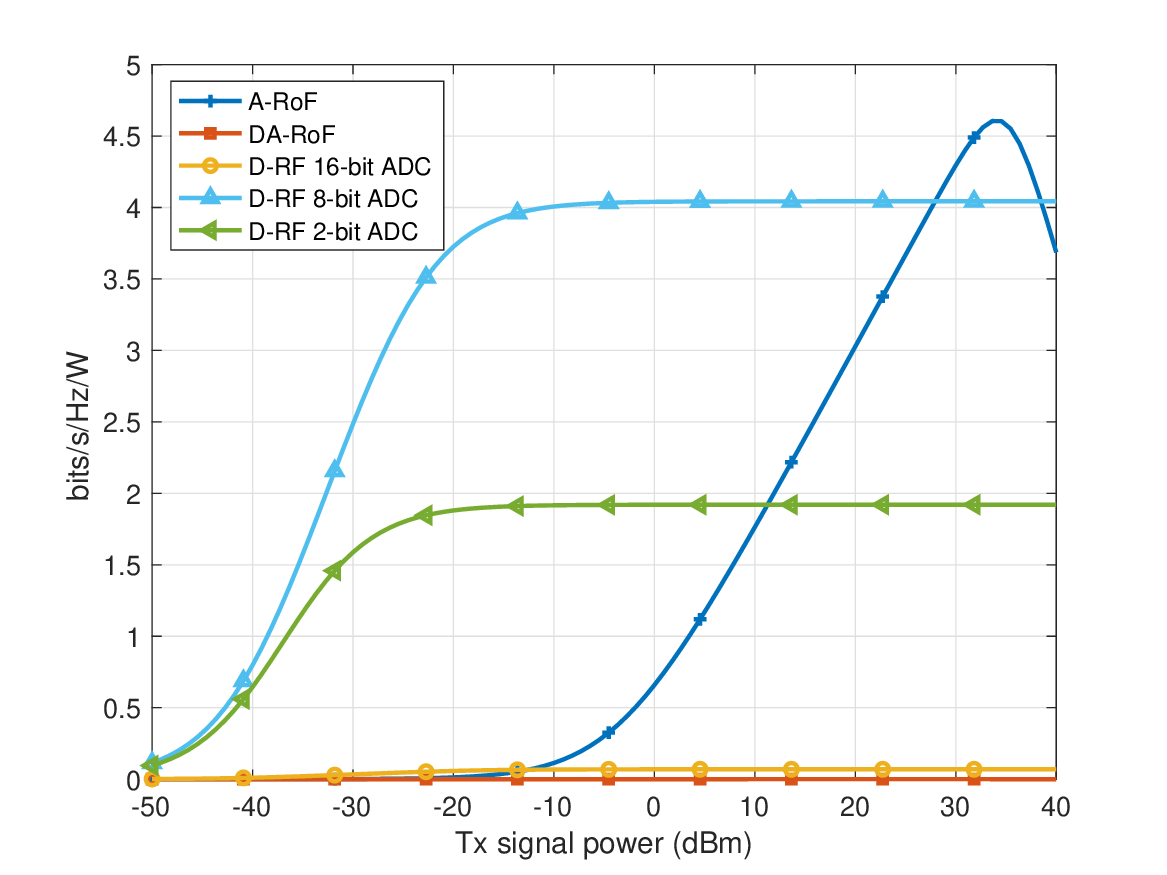}\caption{Power efficiency of all systems. The DA-RoF system has the lowest energy efficiency. In the case of the A-RoF system, the energy efficiency increases, and before its nonlinear point, it can achieve higher efficiency compared to linear low-resolution receivers. \protect\label{fig:Power-spectral-efficiency}}
\end{figure}

\begin{figure}[t]
\centering
\includegraphics[width = \columnwidth]{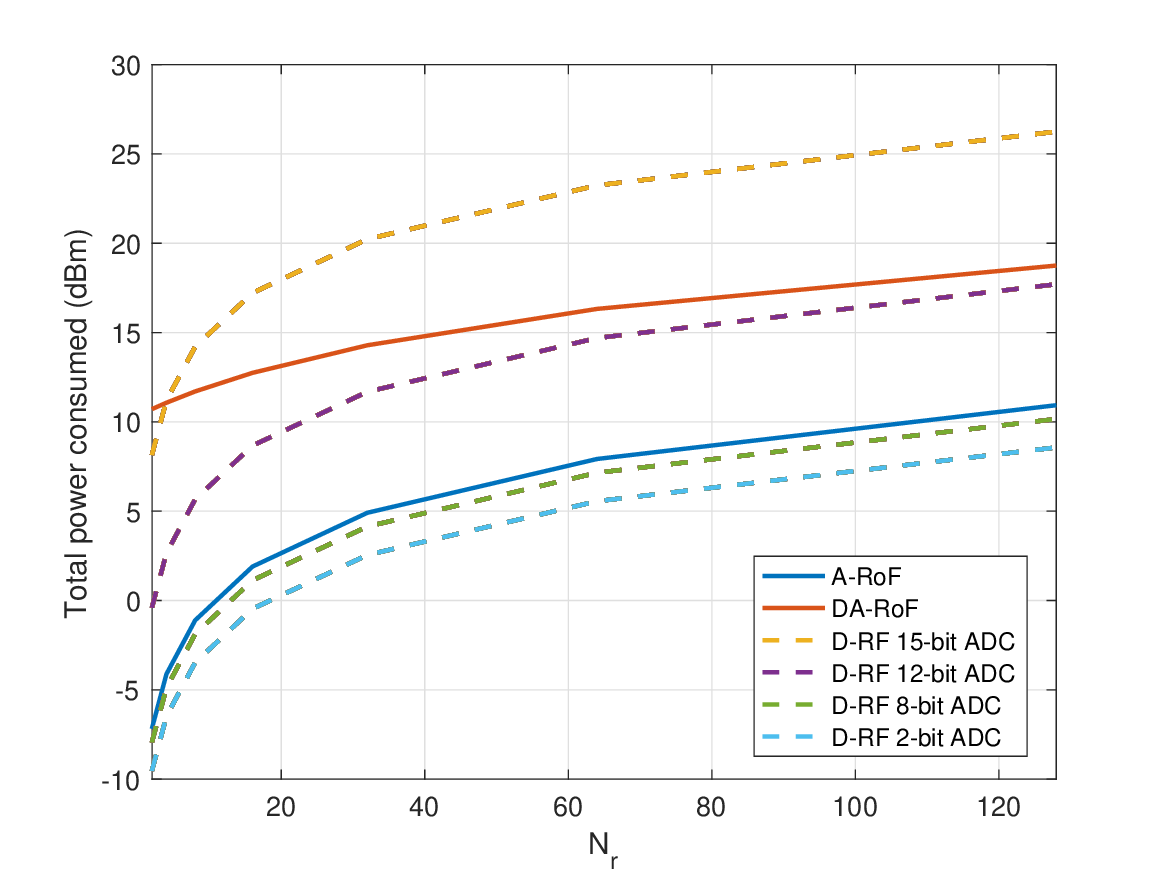}\caption{Comparing total consumed power at the RRH against varying number of receive antennas for RoF systems. \protect\label{fig:PowervsNr}}
\end{figure}

\begin{figure}[t]
\centering
\includegraphics[width = \columnwidth]{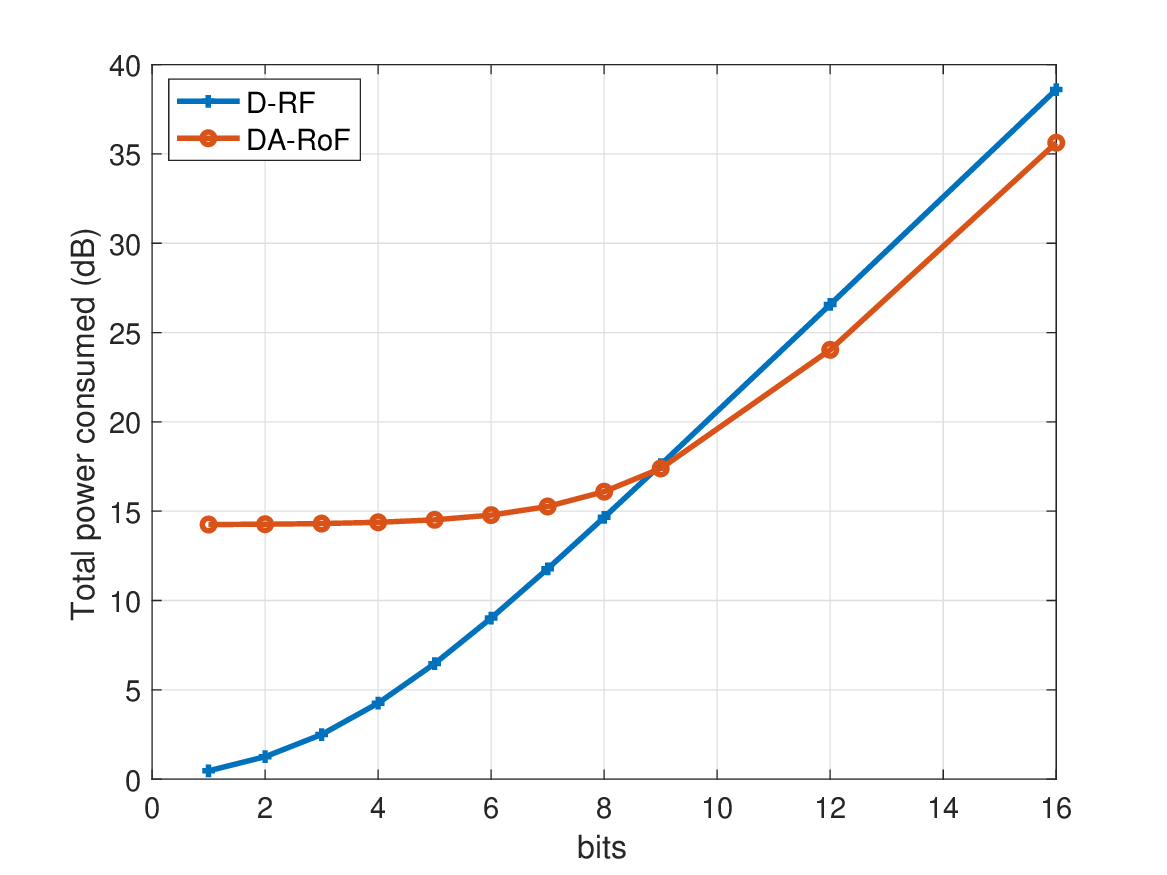}\caption{The power consumption of a DA-RoF receiver and D-RF varies with the number of ADC quantization bits. The total power consumed by the DA-RoF receiver is higher for fewer bits. As the number of bits increases, the power consumption of D-RF receivers surpasses that of the D-RF receiver. \protect\label{fig:D-RoF_power}}
\end{figure}

In this section, we numerically analyze the spectral efficiency and power consumption of RoF systems, taking into account impairments in both the optical and wireless channels. The D-RF system considered in this work assumes that significant loss occurs due to quantization effects. In the RoF systems, to isolate the impact of noise impairments, we assume the wireless channel is frequency-flat. We consider three propagation models for the wireless channel: additive white Gaussian noise (AWGN), line-of-sight (LOS), and Rayleigh fading. The transmitted signal has a center frequency of $2.4$ GHz and a bandwidth of $100$ MHz. The rest of the key system parameters are summarized in Table \ref{table:SysParam}.

The addition of thermal wireless noise and dispersion reduces signal quality over fiber. We observe the effect of the wireless channel with only large-scale fading loss referred to as AWGN wireless channel in Fig. \ref{fig:differentRoFmodels-1}. In Fig. \ref{fig:differentRoFmodels-1}, the optical channels for $N_{\text{r}}=\{4,32,64\}$ are simulated, and the results show that due to transmission loss over the optical channel, the spectral efficiency gain is reduced as expected. The underlying reason is that the electrical noise introduced by the wireless link is scaled by the optical transmission power. Fig. \ref{fig:differentRoFmodels-1} further indicates that as the antenna size increases, so does the spectral efficiency, but the peak analog optical transmission rate shifts toward lower transmitted powers. This is because the total power transmitted over the fiber increases with antenna size, resulting in an increase in the interference between WDM channels. This effect can also be influenced by the distance between the transmitter and receiver, as a greater distance results in more signal degradation. However, here the distance between transmitter and receiver is assumed to be the same.

In contrast, as shown in Fig. \ref{fig:differentRoFmodels-2}, the spectral efficiency of the DA-RoF system decreases more significantly than that of the A-RoF system, due to added quantization noise. While the ADC at RRH has 16-bit resolution, a higher-resolution DAC is also required at BBU to mitigate the effects of quantization loss. At higher powers, the A-RoF system achieves higher spectral efficiency, whereas at lower optical powers, DA-RoF exhibits superior performance. The peak spectral efficiency for DA-RoF systems also occurs at lower transmission power. The results from Fig. \ref{fig:differentRoFmodels-2} indicate that simply increasing the SNR of the received wireless signal will not maximize the system performance. Therefore, minimizing digital components, as demonstrated by the A-RoF system, can lead to more promising results. Additionally, enhanced nonlinear dispersion compensation techniques in the fiber could potentially shift the onset of the nonlinear region to higher power levels, further improving system performance.

To further analyze and compare the spectral efficiencies of A-RoF, DA-RoF, and D-RF, we simulated receivers using an AWGN channel. The first case consists of a wireless receiver aided by an A-RoF system with nonlinear optical impairments. The loss associated with quantization noise is incorporated into the analysis via AQNM, as described in Section \ref{sec:System-model}. We consider high and low-resolution ADCs for electrical D-RF receivers having 32, 16, and 8 bits, respectively. Lastly, we consider DA-RoF, which consists of ADCs and optical elements at the RRH. The loss of information due to ADCs affects the transmission performance and the information being transmitted over the link. The numerical simulations of the spectral efficiency of all the systems are shown in Fig. \ref{fig:differentRoFmodels-2}. The DA-RoF system suffers from larger quantization noise, which is being scaled by the optical impairments. The information loss in A-RoF and DA-RoF is due to the nonlinearities of the fiber, which allow better transmission than low-resolution ADCs. We compared the performance of the A-RoF system with the D-RF system with different ADC resolutions, and we observed that for low-resolution ADCs, the performance degrades significantly as compared to A-RoF, as demonstrated in Fig. \ref{fig:differentRoFmodels-2}. These findings establish RoF systems as a viable option for future deployments of radio access networks, offering higher spectral efficiency and greater flexibility in separating the RRH and BBU.

To observe the effects of wireless fading and path loss on RoF systems, we simulated a wireless channel with LOS and Rayleigh fading. The achieved spectral efficiency under varying channel conditions is shown in Fig. \ref{fig:DifferentFading}. The spectral efficiency of A-RoF follows a similar trend to the wireless system in the linear region. Increasing path loss and fading have a more significant impact on the DA-RoF system compared to the A-RoF system. For DA-RoF, performance degrades more severely with changes in received signal strength, whereas A-RoF exhibits relatively robust performance. These observations reveal that noise in RoF is amplified by dispersion in the received electrical signal, further deteriorating the performance of the optical link.

Next, we compare the transmitted power required by the A-RoF and DA-RoF systems before the interference increases and becomes nonlinear. We keep the optical power constant at the E/O converter in this case. As the number of antennas increases, the number of WDM channels required changes, and the maximum peak spectral efficiency gain becomes flat. For the DA-RoF system, the peak becomes flatter more quickly than for the A-RoF. These results align with our analysis, which indicates that increasing the number of WDM channels will lead to increased nonlinear interference but also provide a gain in received information. The results are summarized in Fig. \ref{fig:differentRoFmodels}. We then vary the transmitter power and the maximum spectral efficiency for a range of antenna numbers. Similarly, increasing the number of antennas reduces the optimal transmission power required, indicating that, according to the use case, the system can be modified to achieve this. The results indicate a trade-off where increasing transmission power reduces spectral efficiency. We then compare the optical power variation, keeping the transmission power constant, with peak spectral efficiency in Fig. \ref{fig:Comparingtheopt}. We observe a similar trend, indicating an optimal power for the optical laser. This allows for the design of RoF systems that maximize gain while minimizing nonlinear interference.

Next, we compare the total RRH power consumption for A-RoF, DA-RoF, and D-RF. The power consumed for analog transmission over fiber increases close to a constant and then grows exponentially as the input optical power increases exponentially, due to the intrinsic properties of the EDFA amplifier. This supports the results in Fig. \ref{fig:Comparingtheopt}, as increasing the optical transmission power further amplifies nonlinear interference. As shown in Fig. \ref{fig:Comparingthepower}, A-RoF consumes less power than both DA-RoF and high-resolution D-RF in most configurations. The A-RoF power consumption of a 20 dBm laser modulator enables it to achieve performance similar to that of an 8-bit ADC receiver. It is also observed that as the input laser power increases, the power consumed by the laser changes, which in turn increases the power consumed by the amplifier, as shown in (\ref{eqn:EDFA}). A short-reach A-RoF link can provide a power-efficient communication method, as the optical interconnect power consumption remains low, thereby minimizing its impact on nonlinear interference.

Moreover, we compare the energy efficiency of RoF systems in Fig. \ref{fig:Power-spectral-efficiency}. The results show the $\text{SE}/P_{\text{tot}}$ values for different receivers, while maintaining a laser power of around 15 dBm. The energy efficiency achieved by A-RoF is highest at around 28 dBm of transmitted power, whereas digital receivers with 2-bit and 8-bit ADCs achieve higher overall energy efficiency. A-RoF, however, consumes power comparable to that of low-resolution receivers and allows for higher spectral efficiency over the fiber link. It performs comparably better than power-consuming systems, such as DA-RoF, making it more advantageous than its digital counterparts. There's another benefit, not explored here, of extending WDM with analog combining at the RRH, which can provide further flexibility. 

Furthermore, we compare power consumption as the number of antennas increases, and the results are summarized in Fig. \ref{fig:PowervsNr}. As the number of antennas increases, the total power consumed by the RRH also increases, as expected. The difference in power consumption between A-RoF and the receiver with an 8-bit ADC is much lower than that between DA-RoF and the receiver with an 8-bit ADC. So, the D-RF-enabled system is not a scalable solution for MIMO or massive MIMO. This result implies that power spectral efficiency decreases with the number of antennas. However, A-RoF consistently outperforms its digital counterparts. Given appropriate nonlinear compensation and higher-order multiplexing techniques, A-RoF can achieve even greater performance gains than D-RF.

Lastly, we compare the increase in power consumption as the number of bits increases for ADC, as shown in Fig.\ref{fig:D-RoF_power}. It compares the increase in power consumption with ADC resolution. Initially, DA-RoF systems consume more power at lower ADC resolutions than D-RF systems. However, as ADC resolution increases, the power consumption of electrical RF receivers in D-RF systems rises significantly, surpassing that of DA-RoF systems. This trend occurs because optical receivers in DA-RoF systems inherently require lower power, offsetting the increased power demands of higher-resolution ADCs. Consequently, DA-RoF can serve as a suitable alternative, enabling additional processing at the RRH, thereby allowing the BBU to be placed remotely. Although DA-RoF consumes substantially more power than purely A-RoF, it effectively supports large-scale MIMO systems. Furthermore, analog-aided RoF systems offer dual advantages: they enable the functional separation of baseband and radio processing, and reduce overall receiver power consumption.

The results here show that A-RoF and DA-RoF receiver performance is ultimately limited by nonlinearities in the optical fronthaul, as launch power or the number of RF chains increases, nonlinear interference dominates and prevents further gains in spectral efficiency. In the linear operating region, however, A-RoF scales efficiently with large antenna arrays and low RRH power, making it well-suited for short-reach, multi-antenna deployments. The Gaussian noise model used here provides a conservative lower bound on performance, with real systems expected to fall between this model and ideal AWGN predictions, consistent with recent studies. More accurate nonlinear-interference models, such as enhanced Gaussian noise formulations, can further tighten these bounds \cite{CarenaEtAlEGNModelNonlinearFiber2014, AmirabadiEtAlClosedFormEGNModelFMF2021, SecondiniEtAlNonlinearityMitigationWDMSystems2019}. Furthermore, in this case, we have assumed that each WDM channel is connected to a separate antenna, which is a baseline assumption. We can also consider a hybrid scenario in which each phased array is supported by WDM \cite{LiEl-HajjarIntelligentAnalogRadioFiber2022}. The system described here can inform future work on channel estimation and beamforming that explicitly incorporates optical impairments. Future work can also leverage analog relay concepts to enhance scalability and efficiency in massive MIMO and extend to cascaded RoF deployments for distributed systems.

\section{Conclusion} \label{sec:Conclusion}

In this work, we quantify the impact of nonlinear interference on analog radio transmission over single-mode fiber. We derived the signal model under the assumption that coherent analog transmission is present, and chromatic dispersion is compensated, enabling us to focus on the noise impairments in A-RoF. We analyzed the spectral efficiency of A-RoF and DA-RoF systems and numerically compared the impact of nonlinear interference. We further extended our analysis to digital radio receiver systems equipped with low-resolution ADCs, accounting for specific impairments at the antenna unit, such as quantization. The numerical results demonstrate that A-RoF offers higher energy efficiency and is power-efficient for large antenna arrays. The DA-RoF system exhibits higher quantization noise because it is amplified by an optical amplifier, but offers higher spectral efficiency at low SNR. The higher spectral efficiency at low transmitted power levels indicates that DA-RoF offers a trade-off in terms of added RRH functionality. A-RoF performance can be further improved through an efficient BBU design, since the optical channel's chromatic dispersion is deterministic and can be digitally compensated. Future work can leverage the deterministic nature of dispersion to design impairment-aware combining at the BBU, and extend the current single-user SIMO setting to multi-user and massive MIMO scenarios with joint optical–wireless resource allocation.

\bibliographystyle{IEEEtran}
\bibliography{IEEEabrv,References,Ref2}


\end{document}